\documentstyle[aps,twocolumn,prl,epsfig,ilja1]{revtex}

\begin{document}

\author{Dragomir Neshev$^1$, Alexander 
	Nepomnyashchy$^{1,2}$, and Yuri S. Kivshar$^1$}

\address{$^1$Nonlinear Physics Group, Research School of Physical 
Sciences and Engineering, The Australian National University
Canberra, ACT 0200, Australia}

\address{$^2$Department of Mathematics and Minerva Center for
Nonlinear Physics of Complex Systems, Technion - Israel Institute of 
Technology, 32000 Haifa, Israel}

\title{Nonlinear Aharonov-Bohm scattering by optical vortices}
\maketitle

\begin{abstract}
We study {\it linear} and {\em nonlinear} wave scattering by an optical 
vortex in a self-defocusing nonlinear Kerr medium. In the linear 
case, we find a splitting of a plane-wave front at the vortex 
proportional to its circulation, similar to what occurs in the 
scattered wave of electrons for the Aharonov-Bohm effect. For larger 
wave amplitudes, we study analytically and numerically the scattering 
of a dark-soliton stripe (a nonlinear analog of a small-amplitude 
wavepacket) by a vortex and observe a significant asymmetry of the 
scattered wave. Subsequently, a wavefront splitting of the scattered 
wave develops into {\em transverse modulational instability},
``unzipping'' the stripe into trains of vortices with opposite charges.
\end{abstract}

It is known that the magnetic vector potential influences the
dynamics of a charged quantum particle even if the magnetic field
vanishes, e.g. when the field is confined to a cylinder into which
electrons can not penetrate. This phenomenon is known these days 
as the Aharonov-Bohm effect \cite{a-b}, and it was shown to have a
more general geometrical framework being directly linked to the wave
front dislocations and geometrical phases \cite{berry}. It was also 
shown to have a classical analog in the scattering of a linear waves 
by a vortex \cite{berry2}. Recently, the study of the interaction of
 a water wave with a vortex  \cite{lund} revealed both similarities 
and differences with the  Aharonov-Bohm effect, allowing to observe 
directly the macroscopic  aspects of the geometrical phases. 

In this Letter we study {\it an optical analog of the Aharonov-Bohm 
effect} - the wave scattering by an optical vortex in a bulk dielectric 
nonlinear medium. We describe, for the first time to our knowledge, 
{\em   nonlinear effects in  the Aharonov-Bohm scattering} including 
both the crossover regime and  strongly nonlinear case. Since our basic equations 
describe  also the dynamics of a superfluid flow in the Bose-Einstein 
condensates \cite{bec}, we expect that our results will be useful for 
the experimental study of the similar effects in the nonlinear matter-wave
physics.

{\it Model.}
We consider the propagation of a monochromatic scalar electric field 
in a bulk nonlinear Kerr-like medium with an intensity-dependent
refractive index, $n=n_0+n_2I$, where $n_2$ is the Kerr-effect
coefficient characterizing a defocusing nonlinearity ($n_2<0$). In the
paraxial approximation, the Maxwell equations can be reduced to the 
nonlinear Schr\"odinger (NLS) equation for the renormalized slowly 
varying amplitude $\psi$, 
\begin{equation}
\label{eq2}
i\frac{\partial \psi}{\partial z} + \frac12\nabla_{\perp}^2 \psi 
	+ (1-|\psi|^2)\psi =0,
\end{equation}
where the propagation coordinate ($z$) and the transverse coordinates
($x,y$) are measured in the units of $k_0|n_2|I_0$ and
$(n_0|n_2|k_0^2I_0)^{1/2}$, respectively. Here $k_0$ is the free-space 
wave number, $n_0$ is the linear refractive index, $\nabla_{\perp}$ is 
the transverse gradient operator, and $I=|E|^2$ is the field intensity
(see details in Ref.~\cite{dark_rev}).

Equation (\ref{eq2}) has a stationary solution for a vortex \cite{dark_rev},
$\psi_0(r)=R_0(r)e^{i\varphi}$, where the real amplitude $R_0(r)$
satisfies the conditions $R_0(0)=0$, and $R_0(r)=1+O(1/r^2)$, as
$r\rightarrow \infty$. To study the wave scattering by the vortex, we
introduce a new function $\Phi(r,z)$, substituting
$\psi(r,z)=\psi_0(r)\Phi(r,z)$ into Eq.~(\ref{eq2}) and taking into
account the lowest-order terms in $r^{-1}$. The corresponding equation
for $\Phi(r,z)$ has the form,
\[
i\left( \frac{\partial\Phi}{\partial z}
	+\frac{({\vec e}_\varphi\cdot\nabla_{\perp})\Phi}{r} \right) 
	+\frac12\nabla^2_{\perp}\Phi + (1-|\Phi|^2)\Phi =0,
\]
and it describes the advection of the field $\Phi(r,z)$ by the vortex 
phase gradient field ${\vec e}_\varphi/r$.

{\it Linear limit: the Aharonov-Bohm scattering.} 
First, we analyze the scattering of a small-amplitude wave on a vortex.
This problem was studied earlier in the context of the superfluid 
hydrodynamics \cite{pismen}. Herewith, we develop a different approach 
that allows a simple further extension to the nonlinear regime.

Let us consider a plane wavefront with the coordinate $x=X(y)$ that
moves in the $x$-direction with the velocity $v$ [see Fig.~1(a)], and
is advected simultaneously by the vortex velocity field
$u^{(0)}(y,z)=-(y/r^2)=-y/(y^2+v^2z^2)$. If the wavefront motion
starts at $z=z_{\rm in}<0$ and the point $z=0$ corresponds to 
the moment when the unperturbed line crosses the vortex [see Fig.~1(c)], 
the shape of the line in the $y$-direction, $h^{(0)}(y,z)$, can be 
found as (for $|y|\gg 1$):
\begin{equation}
\label{eq4}
   h^{(0)}(y,z)  =
   -\frac{\mbox{sign}(y)}{v}\left(\tan^{-1}\frac{vz}{|y|} -
		\tan^{-1}\frac{v z_{\rm in}}{|y|}\right). 
\end{equation}
Thus, the total deformation of the wave front can be calculated as follows
(for $v|z|,v |z_{\rm in}| \gg |y|\gg 1$):
\begin{equation}
\label{eq5}
h^{(0)}(y,z)= -\frac{\mbox{sign}(y)}{v}\pi.
\end{equation}
Equation (\ref{eq5}) coincides with the result given by Eq.~(4.73) of 
Ref.~\cite{pismen} provided $v$ has a meaning of the sound speed. 
It describes a splitting of the wavefront by the 
amount $\Delta x=-(2\pi/v)$, predicted by the Aharonov-Bohm effect for 
the linear waves scattered by a vortex.

\vspace{-3mm}
\begin{figure}
\setlength{\epsfxsize}{3.2in}
\centerline{\epsfbox{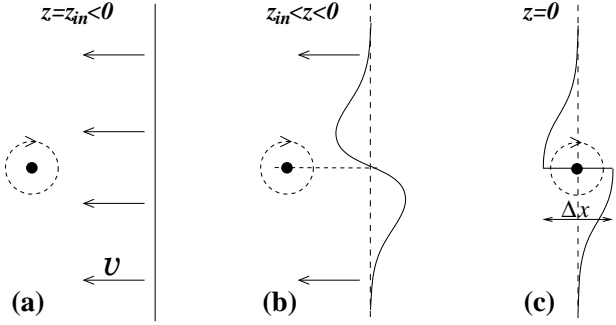}}
\vspace{1mm}
\caption{Sketch of the wavefront deformation and splitting at different 
distances from the vortex.}
\end{figure}
\vspace{-2mm}

{\it Nonlinear regime: Deformation of a dark-soliton stripe.}
In the nonlinear regime of the vortex-wave scattering,
the steady-state propagation of the wave packet in the presence of 
nonlinearity is possible in the form of a solitary wave when 
nonlinear effects are balanced by the wave dispersion \cite{note}. 
In the problem under consideration, such a solitary wave corresponds 
to a localized intensity dip on a constant plane-wave background, a 
{\it spatial dark-soliton stripe} \cite{dark_rev}.
The corresponding {\em nonlinear scattering problem} is rather 
complicated, and we restrict our study by the region of large 
distances from the vortex core, $r=O(\epsilon^{-1})$, where 
$\epsilon\ll 1$ is a formal small parameter of the asymptotic 
analysis we develop below. 

We assume that, under the action of the vortex field, a dark-soliton 
stripe moves with the velocity $v=O(1)$ and the 
coordinate of its center gets shifted, in respect to its unperturbed 
motion $x=vz$, by the amount $h(\epsilon y, \epsilon z)=O(1)$. 
Introducing new variables
$Y=\epsilon y$, $Z=\epsilon z$, $\xi=x-vz-h(Y,Z)$, we look for 
the solution in the asymptotic form,
$\Phi=\sum_{n=0}^\infty\epsilon^n\Phi^{(n)}$. In the zero order in
$\epsilon$, we find 
$$\frac12\frac{\partial^2\Phi^{(0)}}{\partial\xi^2}
	- iv\frac{\partial\Phi^{(0)}}{\partial\xi}
	+ (1-|\Phi^{(0)}|^2)\Phi^{(0)}=0 ,$$
that has a well-known solution for a dark soliton \cite{dark_rev}
\begin{equation}
\label{eq7}
\Phi^{(0)}=[\sqrt{1-v^2}\tanh(\sqrt{1-v^2}\xi)+iv]e^{i\theta} .
\end{equation}
In the first order in $\epsilon$, we obtain the equation
\begin{equation}
\label{eq8}
 \begin{array}{lll}
   {\displaystyle \frac12\frac{\partial^2\Phi^{(1)}}{\partial\xi^2} }
	&-& 
   {\displaystyle iv\frac{\partial\Phi^{(1)}}{\partial\xi}
	+ (1-2|\Phi^{(0)}|^2) \Phi^{(1)} } \\
	&-& 
   {\displaystyle \Phi^{(0)\,2}\Phi^{\ast\, (1)} =
	iV(Y,Z)\frac{\partial\Phi^{(0)}}{\partial\xi} },
 \end{array}
\end{equation}
where
\begin{equation}
\label{eq9}
V(Y,Z)=\frac{\partial h}{\partial Z}+\frac{Y}{Y^2+v^2Z^2} .
\end{equation}
A solution of Eq.~(\ref{eq8}) can be found as
$\Phi^{(1)}=V(Y,Z)(\partial\Phi^{(0)}/\partial v)$. Result 
(\ref{eq9}), being expressed as $h_Z = V - Y(Y^2 + v^2Z^2)^{-1}$, 
has a simple physical meaning: a local change of the stripe 
coordinate, $h_Z$, is a sum of its local velocity, $V$, and 
a correction produced by the vortex.

In the second order in $\epsilon$, we obtain an equation for 
$\Phi^{(2)}$ similar to Eq.~(\ref{eq8}) with different r.h.s. 
terms. The equation governing the evolution of the 
dark-soliton-stripe parameters can be obtained from the 
orthogonality condition of the r.h.s. of the equation 
for $\Phi^{(1)}$ to the solution of the homogeneous problem,
$\Phi^{(1)}=\partial\Phi^{(0)}/\partial\xi$. Taking into account the
parity properties and assuming the normalization 
$\mbox{Im}(\partial\Phi^{(0)}/\partial v)=1$, we derive the equation
\begin{equation}
\label{eq10}
\frac{\partial V}{\partial Z}
	+\lambda^2(v)\frac{\partial^2h}{\partial Y^2}=0,
\end{equation}
where $\lambda^2(v)= \frac{1}{3} (1-v^2)$. System of 
Eqs.~(\ref{eq9}), (\ref{eq10}) describes the evolution of the 
stripe parameters in the presence of a vortex field, and it can 
be rewritten in the original variables ($y,z$), if we define 
$\tilde u=\epsilon V$, and $h_Z=\tilde u+u^{(0)}$, as follows
\begin{equation}
\label{eq11}
\frac{\partial^2\tilde u}{\partial z^2}
	+\lambda^2\frac{\partial^2 \tilde u}{\partial y^2}=
	-\lambda^2\frac{\partial^2 u^{(0)}}{\partial y^2} .
\end{equation}

{\it Nonlinear wavefront splitting.} 
The Cauchy problem for the elliptic equation (\ref{eq11}) is ill-posed, 
hence its solutions are meaningful only if the generation of 
short-wavelength harmonics is not essential. To solve Eq.~(\ref{eq11}), 
first we apply the Fourier transformation and obtain the equation
\begin{equation}
\label{eq12}
\frac{d^2u_k}{dz^2}-k^2\lambda^2u_k=i\pi k^2\lambda^2\mbox{sign}(k)
	e^{-|k|v|z|},
\end{equation}
where $u_k(z)$ is the Fourier transform of the field $\tilde u(y,z)$.
Except the special case $\lambda=v$, the solution of Eq.~(\ref{eq12}) 
can be found in a general form
\begin{equation}
\label{eq13}
u_k=\left\{\matrix{
 {\displaystyle
   A_k^-e^{k\lambda z}+B_k^- e^{-k\lambda z}+
     a_ke^{|k|vz}},&z<0,\cr
 {\displaystyle
   A_k^+e^{k\lambda z}+B_k^+ e^{-k\lambda z}+ a_ke^{-|k|vz}},&z>0, }
\right.
\end{equation}
where $a_k=i\pi\lambda^2(v^2-\lambda^2)^{-1}\mbox{sign}(k)$, and the 
values of the amplitudes $A_k^\pm$ and $B_k^\pm$ are 
determined from the continuity conditions,
$$A_k^+ -A_k^- =-(B_k^+ - B_k^-)=\frac{i\pi\lambda v}{(v^2-\lambda^2)}.$$

First, we consider the region $z<0$. If the stripe instability is weak, 
i.e. $\lambda<v$, the third term in Eq.~(\ref{eq13}) dominates, and we find
the result: $\tilde u(y,z)\approx [\lambda^2/(v^2-\lambda^2)]u^{(0)}(y,z)$.
Thus, the deformation of a dark-soliton stripe due to its
interaction with the vortex field can be presented in the form,
\begin{equation}
\label{eq14}
h(y,z)=\frac{v^2}{(v^2-\lambda^2)}h^{(0)}(y,z), \;\;\; z<0,
\end{equation}
where $h^{(0)}(y,z)$ is defined by Eq.~(\ref{eq4}) of the linear theory.
Therefore, the result (\ref{eq14}) describes the so-called {\it nonlinear
Aharonov-Bohm scattering} on the vortex; it is stronger than its linear
analog by the factor $v^2/(v^2-\lambda^2)$, 
provided $\lambda<v$ as $v>1/2$.

The general solution (\ref{eq13}) describes {\it a growth of instability 
of a dark-soliton stripe}. If this instability is rather strong 
(i.e. $\lambda>v$), the term with $A_k^-$ becomes important, and the 
amplitude coefficients $A_k^-$ and $B_k^-$ should be determined from 
the initial conditions at $z=z_{\rm in}$. If 
$h(y, z_{\rm in})=h_z(y, z_{\rm in})=0$, in the presence of a vortex
the function $u_k$ grows {\it linearly}, for small $z+z_{\rm in}$, 
and {\it exponentially}, for large $z+ z_{\rm in}$. This type of the 
stripe evolution was earlier discussed in Ref.~\cite{ol} for the 
special case $v=0$.

More importantly, we can find {\it an exact non-singular
analytical solution} of Eq.~(\ref{eq13}) for any $v$ and $\lambda$.
Assuming $z=z_{\rm in}<0$ and selecting the initial conditions in the form
$\tilde u(y,z_{\rm in})=\tilde u_z(y,z_{\rm in})=0$, $h(y,z_{\rm in})=0$,
$h_z(y,z_{\rm in})=u^{(0)}(y,z_{\rm in})$, we solve Eq.~(\ref{eq12})
and find the dark-soliton deformation as following
\begin{equation}
\label{eqA}
 \begin{array}{lll}
   h(y,z)
	&=&
   {\displaystyle \left[ \frac{1}{2(v+\lambda)}\tan^{-1}\left(
     \frac{-\lambda z+(v+\lambda)z_{\rm in}}{|y|}\right)\right. } \\
	&+& 
   {\displaystyle \frac{1}{2(v-\lambda)}\tan^{-1}\left(
	\frac{\lambda z+(v-\lambda)z_{\rm in}}{|y|}\right) } \\
	&-&
   {\displaystyle 
      \left.\frac{v}{v^2-\lambda^2}\tan^{-1}\left(\frac{vz}{|y|}\right) 
		\right]\mbox{sign}(y) } .
 \end{array}
\end{equation}
Exact analytical solution (\ref{eqA}) is more general than the result
(\ref{eq14}), and it transforms into Eq.~(\ref{eq14}) for 
$\left\vert (v - \lambda)z_{in}/y\right\vert \gg 1$. Solution (\ref{eqA})
remains valid even for $v\rightarrow 0$, i.e. for deformations 
of a black soliton.
 
In the region $z>0$, the exponentially growing term $\sim A_k^+$ of 
Eq.~(\ref{eq13}) dominates, and the corresponding divergence at large 
$k$ is a direct manifestation of the ill-posed problem (\ref{eq11}). 
It can be regularized by the higher-order term 
$\sim \beta h_{zyy}$, which appears in the equation 
$h_{zz}+\lambda^2 h_{yy}=u_z^{(0)}$ due to the effect of radiation 
taken into account in the asymptotic expansion (see details of such a 
technique in Ref.~\cite{review}). This term makes the linear problem 
(\ref{eq11}) well-posed, but it appears with additional {\it nonlinear 
corrections} that all describe the dynamics of the transverse 
instability of the dark-soliton stripe in the presence of a vortex 
field. The asymptotic analysis becomes very complicated and, in order 
to analyze the long-term dynamics of the nonlinear wave scattering, 
bellow we carry out numerical simulations.

{\it Numerical simulations} have been conducted by means of a 
standard beam propagation method, solving Eq.~(\ref{eq2}) over a 
numerical grid of $1024\times 1024$ points. As initial conditions in
the nonlinear problem, we use a superposition of a vortex and 
a dark-soliton stripe: $\psi(x,y,0)=\psi_0(r)\Phi(x,0)$, where 
$\psi_0(r)=R_0(r)e^{i\varphi}$ and $R_0(r)$ describes the 
stationary solution for an optical vortex soliton found numerically 
by a shooting technique. The function 
$\Phi(x,0)$ describes a dark-soliton stripe with an initial offset 
from the position of the vortex ($x_0$), and it is given by Eq.~(\ref{eq7}). 
To ensure unperturbed propagation of an optical beam over long 
distances, the initial function $\psi(x,y,0)$ was superimposed on a 
super-Gaussian-beam background with HWHM more than 40 times larger 
than the characteristic scale of the localized structures
and interaction domain. 

\vspace{-1mm}
\begin{figure}
\setlength{\epsfxsize}{3.4in}
\centerline{\epsfbox{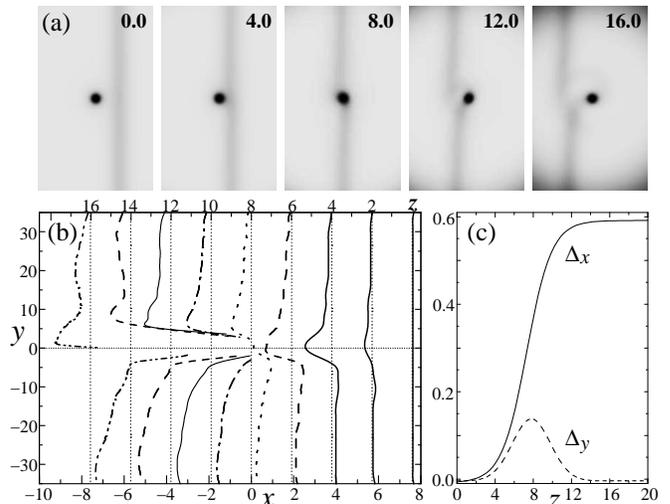}}
\caption{(a) Snapshots of the interaction of an optical 
vortex with a dark-soliton stripe ($v^2=0.9$). Initially, the stripe 
is shifted at $x_0=7.59$ from the vortex and, if propagating 
alone, it reaches the origin at $z=8$. For better visualization,
$16\%$ of the computational domain are shown, and the 
contrast of the images was adjusted. (b) Asymmetric stripe deformation 
at different propagation distances. (c) Vortex horizontal ($\Delta_x$) 
and vertical ($\Delta_y$) shift from its initial position.}
\end{figure}
\vspace{-2mm}

In our simulations, we vary the contrast of the dark-soliton stripe 
(i.e. its transverse velocity), and monitor the stripe deformation 
for different propagation distances. First of all, we investigate the 
scattering of a small-amplitude dark-soliton stripe by an optical 
vortex, which should correspond to the scattering of a linear 
wavepacket associated with the Aharonov-Bohm effect. In Fig.~2(a), we 
show the snapshots of the scattering process for the dark-soliton 
stripe with the transverse velocity $v^2=0.9$, at different propagation 
distances $z$ (marked on the figure). Below in Fig.~2(b,c), we show the 
details of the stripe deformation (first {\it symmetric} but then 
{\it asymmetric}) during the scattering, and the vortex shift $\Delta$
induced by the stripe scattering \cite{note2}. 

To investigate the strong nonlinear regime of the vortex-stripe scattering 
we consider different values of the stripe amplitude by changing its 
velocity and, therefore, contrast. The results of these calculations are 
summarized in Fig.~3, where we show the results of the scattering process 
at the propagation distance $z=16$, similar to the corresponding plot in 
Fig.~2(a). We notice that the transverse modulational instability of the 
stripe starts developing for intermediate dark-stripe contrasts, and it 
is accompanied by a subsequent formation of mixed (edge-screw) and, later, 
screw phase dislocations. Consequently, the stripe breaks-up into vortices 
of opposite charges in a way that resembles the dynamics and formation 
of vortex streets in the hydrodynamics. This phenomenon can be 
understood as an effective ``unzipping'' of the dark-soliton stripe by 
the vortex, that is a manifestation of strongly nonlinear regime in the 
Aharonov-Bohm scattering, a remarkable nonlinear effect. 

\vspace{-2mm}
\begin{figure}
\setlength{\epsfxsize}{3.2in}
\centerline{\epsfbox{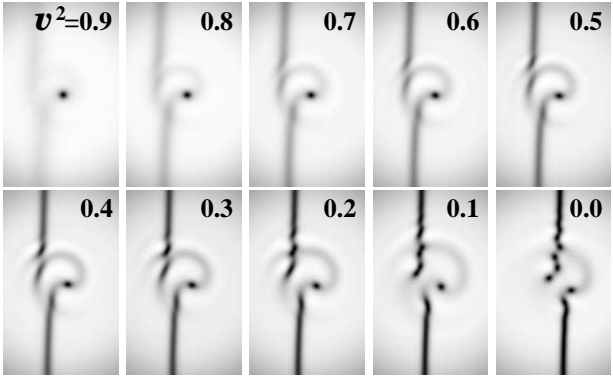}}
\vspace{1mm}
\caption{Snapshots of a dark-soliton stripe scattered by an
optical vortex soliton, after propagation $z=16$ and at different 
initial velocities $v^2$. The initial offset of the stripe is set such 
that if alone the stripe reaches the origin at $z=8$, except for the case 
$v=0$ where $x_0=2.5$.}
\end{figure}
\vspace{-1mm}

\vspace{-2mm}
\begin{figure}
\setlength{\epsfxsize}{3.2in}
\centerline{\epsfbox{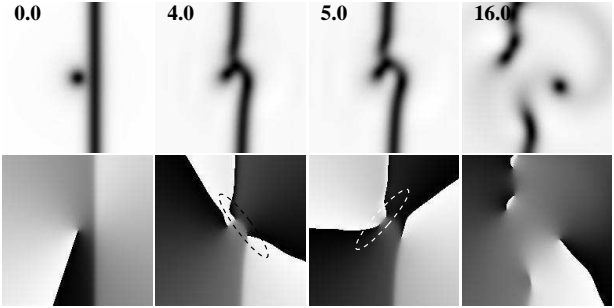}}
\vspace{1mm}
\caption{Vortex pair creation and annihilation during the scattering of the
stripe ($v^2=0.1$), that gives rise to an effectively large vortex shift.}
\end{figure}
\vspace{-1mm}

Additionally, we notice that the primary vortex undergoes a large shift 
from its initial position after the scattering. In order to understand this 
effect, we have studied this process in details (see  Fig.~4) and have observed
a novel type of interaction associated with the creation/annihilation process 
of a vortex pair. Indeed, during the interaction, a part of the stripe closest 
to the vortex breaks up creating a pair of vortices with opposite charges
[see Fig.~4, $z=4$]. Consequently, one of the new vortices annihilates with 
the primary vortex [Fig.~4, $z=5$], substituting it by a new vortex of the 
generated pair. This appears as an overall effect of a large vortex shift 
[Fig.~4, $z=16$].

{\it Conclusions.}
In this Letter, we have studied {\em an optical analog of the 
Aharonov-Bohm scattering} that allows illustrating, at a macroscopic 
level, one of the fundamental quantum mechanical phenomena. Moreover, 
since each specific problem on the geometrical phases has its own 
peculiarities, in optics we are able to study, for the first time to
our knowledge, both weakly and strongly {\em nonlinear 
effects} in the vortex-induced wave scattering characterized by a 
significant change in the symmetry of the scattered wave and the 
subsequent stripe instability, unzipping and break-up. Here we have
analyzed the NLS model in optics since the experimental verification of the
obtained results is a subject of our following interest. However the same model
is known to appear in the macroscopic dynamics of the Bose-Einstein 
condensates, so that our analysis applies to that case as well. We hope 
our theoretical results can be further verified experimentally, in both 
nonlinear optics and the physics of ultracold condensed gases.

The work was supported by the Performance and Planning Fund of the 
Institute of Advanced Studies and the Australian Photonics Cooperative 
Research Centre. A.A.N. acknowledges a support from the Fund for the 
Promotion of Research at the Technion, and a travel grant of the 
Israeli-Australian Theeman Foundation. The authors thank M. Berry, 
D. Pelinovski, L. Pismen, E.B. Sonin, and M. Soskin for 
their interest to this work, and V. Tikhonenko, J. Christou, and 
B. Luther-Davies for collaboration at the initial stage of this project.


\begin{references}
\bibitem{note} As a matter of fact, Eq.~(\ref{eq5}) is valid in the 
long-wavelength limit when the wave dispersion vanishes, and all linear 
waves propagate with the same (sound speed) velocity.
\bibitem{note2} Far from the vortex, the numerical results are described by 
Eq. (12). For example, at $z_{\rm in} = -16$ and $v^2 =0.9$, we have
 $\Delta x (-4) = 1.36$, that agrees well with the numerical data, $\Delta x(-4) = 1.41 \pm 0.12$.
\end{references}
\end{document}